\documentstyle[aasms4]{article}

\begin{document}

\title{RELATIVISTIC THERMAL BREMSSTRAHLUNG GAUNT 
FACTOR FOR THE INTRACLUSTER PLASMA. II. HEAVY ELEMENTS}

\author{SATOSHI NOZAWA\altaffilmark{1}}

\affil{Josai Junior College for Women, 1-1 Keyakidai, Sakado-shi, Saitama, 350-0295, Japan}

\author{NAOKI ITOH\altaffilmark{2}}

\affil{Department of Physics, Sophia University, 7-1 Kioi-cho, Chiyoda-ku, Tokyo, 102-8554, Japan}

\centerline{AND}

\author{YASUHARU KOHYAMA\altaffilmark{3}}

\affil{Fuji Research Institute Corporation, 2-3 Kanda-Nishiki-cho, Chiyoda-ku, Tokyo, 101-8443, Japan}

\altaffiltext{1}{snozawa@venus.josai.ac.jp}
\altaffiltext{2}{n\_itoh@hoffman.cc.sophia.ac.jp}
\altaffiltext{3}{kohyama@crab.fuji-ric.co.jp}

\begin{abstract}

  We calculate the relativistic thermal bremsstrahlung Gaunt factor for the high-temperature plasma which exists in clusters of galaxies.  We calculate the Gaunt factor by employing the Bethe-Heitler cross section corrected by the Elwert factor.  The calculations in this paper are made for the fully ionized plasma for the following cases: $Z$ = 10 (Ne), 12 (Mg), 14 (Si), 16 (S), 26 (Fe).  We also calculate the Gaunt factor by using the Coulomb-distorted wave functions for nonrelativistic electrons following the method of Karzas and Latter.  By comparing the Gaunt factors calculated by these two different methods, we carefully assess the accuracy of the calculation.  We present the numerical results in the form of tables.

\end{abstract}


\section{INTRODUCTION}

  High-temperature plasmas exist in the clusters of galaxies (Arnaud et al. 1994; Markevitch et al. 1994; Markevitch et al. 1996; Holzapfel et al. 1997).  Some clusters have extremely high-temperature electrons,  $k_{B} T_{e}$ = $10 \sim 15$keV.  Relativistic expressions for the thermal bremsstrahlung emissivity have been discussed by many authors (Gould 1980; Rephaeli \& Yankovitch 1997).  However, the relativistic expressions have been so far derived by power-series expansions.

  Very recently the present authors (Nozawa, Itoh, \& Kohyama 1998) have calculated the relativistic thermal bremsstrahlung Gaunt factor for the intracluster plasma by using the Bethe--Heitler (1934) cross section corrected by the Elwert (1939) factor for the following cases: $Z$ = 1 (H), 2 (He), 6 (C), 7 (N), 8 (O).  They have also calculated the bremsstrahlung Gaunt factor by using the Coulomb--distorted wave functions for nonrelativistic electrons following the method of Karzas \& Latter (1961).  They have thereby assessed the accuracy of the calculations.  Their method of the calculation closely followed the work on the calculation of the inverse thermal bremsstrahlung by two of the present authors and their collaborators (Itoh, Nakagawa, \& Kohyama 1985; Nakagawa, Kohyama, \& Itoh 1987; Itoh, Kojo, \& Nakagawa 1990; Itoh et al. 1991; Itoh et al. 1997).

  In the present paper we will extend the calculation to heavier elements: $Z$ = 10 (Ne), 12 (Mg), 14 (Si), 16 (S), 26 (Fe).  For the sake of completeness, we will restate the formalism in this paper.

  The present paper is organized as follows.  We will give formulations for the calculation of the relativistic thermal bremsstrahlung Gaunt factor in $\S$ 2.  The numerical results will be presented in $\S$ 3.  We will discuss the results and give concluding remarks in $\S$ 4.

\section{FORMULATION}

  In this paper we are concerned with the calculation of the relativistic thermal bremsstrahlung Gaunt factor for a high-temperature, low-density plasma which is relevant to the hot gas in the clusters of galaxies.  We will use the accurate relativistic cross section.  We will neglect the effects of screening and ionic correlation, which are expected to be small for a high-temperature, low-density plasma which is under investigation.

  The cross section for the bremsstrahlung is related to (and can be easily obtained by the principle of detailed balancing from) the inverse bremsstrahlung cross section which is shown in Itoh, Nakagawa, \& Kohyama (1985).  In the last three decades there appeared a great deal of theoretical work concerning the accurate relativistic calculation of the bremsstrahlung cross section (Elwert \& Haug 1969; Tseng \& Pratt 1971; Pratt \& Tseng 1975; Lee et al. 1976).  Elwert \& Haug (1969) and Pratt \& Tseng (1975) have confirmed that the Bethe-Heitler (1934) cross section corrected by the Elwert (1939) factor gives excellent results for ions with small atomic number $Z_{j}$.

  The relativistic cross section for the bremsstrahlung is written (following the notation of the original Bethe-Heitler paper as closely as possible) as
\begin{eqnarray}
\sigma & = & \alpha \, Z_{j}^{2} \, r_{0}^{2} \, \frac{p_{f}}{p_{i}} \, \frac{d \omega}{\omega} \, \frac{a_{f}}{a_{i}} \, \frac{1 - {\rm exp} ( - 2 \pi a_{i})}{1 - {\rm exp} ( - 2 \pi a_{f})}  \nonumber  \\
& \times & \left\{ \, \frac{4}{3} \, - \, 2 E_{f} E_{i} \frac{ p_{f}^{2} \, + \, p_{i}^{2}}{p_{f}^{2} p_{i}^{2} c^{2}} \, + \, m^{2}c^{2} \left[ \,
\frac{\beta_{f} E_{i}}{p_{f}^{3}c} \, + \, \frac{\beta_{i} E_{f}}{p_{i}^{3}c} \, - \, \frac{\beta_{f} \beta_{i}}{p_{f} p_{i}} \right] \right.  \nonumber  \\
& & \, + \, L \left[ \, \frac{8}{3} \, \frac{E_{f} E_{i}}{p_{f} p_{i} c^{2}} \, + \, \frac{\hbar^{2} \omega^{2}}{p_{f}^{3} p_{i}^{3} c^{6}} \, \left( E_{f}^{2} E_{i}^{2} \, + \, p_{f}^{2} p_{i}^{2} c^{4} \right) \right.  \nonumber  \\
& & \, + \, \left. \left. \frac{m^{2}c^{2} \hbar \omega}{2 p_{f} p_{i}} \left( \frac{E_{f} E_{i} + p_{i}^{2}c^{2}}{p_{i}^{3} c^{3}} \beta_{i} \, - \, \frac{E_{f} E_{i} + p_{f}^{2}c^{2}}{p_{f}^{3} c^{3}} \beta_{f} \, + \, \frac{2 \hbar \omega E_{f} E_{i}}{p_{f}^{2} p_{i}^{2} c^{4}} \right) \, \right] \, \right\}  \, ,
\end{eqnarray}
\begin{eqnarray}
a_{f} & \equiv & \frac{\alpha \, Z_{j} \, E_{f}}{p_{f}c} \, , \, \, \, \, \, 
a_{i} \, \equiv \, \frac{\alpha \, Z_{j} \, E_{i}}{p_{i}c} \, ,  \\
\beta_{f} & \equiv & 2 \, \, {\rm ln} \, \frac{E_{f} \, + \, p_{f} c}{mc^{2}}  \, , \, \, \, \, \, \beta_{i} \, \equiv \, 2 \, \, {\rm ln} \, \frac{E_{i} \, + \, p_{i} c}{mc^{2}}  \, ,   \\
L & \equiv & 2 \, \, {\rm ln} \, \frac{E_{f} E_{i} \, + \, p_{f} p_{i} c^{2} \, - \, m^{2}c^{4}}{mc^{2} \hbar \omega}  \, ,  \\
E_{i} & = & E_{f} \, + \, \hbar \omega \, .
\end{eqnarray}
In the above, $\alpha$ is the fine-structure constant, $r_{0}$ is the classical electron radius, $\omega$ is the angular frequency of the emitted photon, $p_{i}$ is the initial momentum of the electron, $p_{f}$ is the final momentum of the electron, $E_{i}$ is the initial energy of the electron, $E_{f}$ is the final energy of the electron.

  We will calculate the bremsstrahlung emissivity
\begin{equation}
W(\omega) \, d \omega \, = \, \hbar \omega \, n_{e} \, n_{j} \, v_{i} \, \sigma \, = \, \hbar \omega \, n_{e} \, \frac{p_{i}c^{2}}{E_{i}} \, n_{j} \, \sigma \, , 
\end{equation}
where $n_{e}$ is the number density of electrons and $n_{j}$ is the number density of ions of charge $Z_{j}$.  Then we will average the bremsstrahlung emissivity over a distribution of electrons taking into account the Pauli blocking of the final electron state
\begin{eqnarray}
< W(\omega) > d \omega & = & \frac{ \displaystyle{ \int W(\omega) \, d \omega \, f(E_{i}) [ 1 - f(E_{f})] \, d^{3} p_{i}}}{ \displaystyle{\int f(E_{i}) \, d^{3} p_{i}}}  \, ,  \\
f(E_{i}) & \equiv & \left\{ {\rm exp} \left[ (E_{i} - \mu)/k_{B} T \right] \, + \, 1 \right\}^{-1}  \, ,  \\
\int f(E_{i}) \, d^{3} p_{i} & = & 4 \pi m^{3} c^{3} \, G_{0}^{-}(\lambda, \nu)    \, ,  \\
G_{0}^{-}(\lambda, \nu) & \equiv & \lambda^{3} \, \int_{\lambda^{-1}}^{\infty} \frac{ x (x^{2} - \lambda^{-2})^{1/2}}{1 \, + \, e^{x - \nu}} \, d x  \, ,  \\
\lambda & \equiv &  \frac{k_{B}T}{m c^{2}} \, = \, \frac{T}{5.930 \times 10^{9} {\rm K}}  \,  ,  \\
\nu  & = & \frac{\mu}{k_{B} T}   \,  ,
\end{eqnarray}
$\mu$ being the electron chemical potential including the rest mass.  For the extreme non-degeneracy ($-\eta \gg 1$) and the nonrelativistic temperature ($\lambda \ll 1$), the chemical potential $\mu$ is related to the electron number density $n_{e}$ and the temperature $T$ through the relationship (relativistic Maxwellian distribution)
\begin{eqnarray}
\eta & \equiv & \frac{\mu \, - \, m c^{2}}{k_{B} T}  \nonumber \\
& = & {\rm ln} \, \left\{ \frac{1}{2} n_{e} \left(\frac{2 \pi \hbar^{2}}{m k_{B} T} \right)^{3/2} \, \left[ 1 \, + \, \frac{15}{8} \frac{k_{B} T}{m c^{2}} \, + \, \frac{105}{128} \left( \frac{k_{B} T}{m c^{2}} \right)^{2} \right]^{-1} \, \right\}  \nonumber  \\
& = & {\rm ln} \, \left\{ 4.535 \times 10^{-31} \left[n_{e}({\rm cm^{-3}}) \right] \lambda^{-3/2} \, \left(1 \, + \, \frac{15}{8} \lambda \, + \, \frac{105}{128} \lambda^{2} \right)^{-1} \right\}  \, .
\end{eqnarray}
The degeneracy parameter $\eta$ is generally related to the mass density and temperature through the relationship
\begin{eqnarray}
\frac{\rho}{2} \left( 1 \, + \, \frac{0.992 X}{1.008} \right) & = &
\frac{n_{e}}{N_{A}} \, = \, \frac{1}{ N_{A}} \frac{2}{( 2\pi \hbar)^{3}} \int f(E_{i}) \, d^{3} p_{i} \,  \nonumber \\
 & = & 2.922 \times 10^{6} \, G_{0}^{-}( \lambda, \eta + \lambda^{-1} ) \, .
\end{eqnarray}
In equation (14) $\rho$ is measured in units of gcm$^{-3}$, $X$ is the mass fraction of hydrogen, $n_{e}$ is the electron number density in units of cm$^{-3}$, and $N_{A}$ is Avogadro's number.  The reader might wonder the reason why we retain the general Fermi-Dirac distribution function rather than we use the relativisitic Maxwellian distribution function from the outset.  The reason is twofold.  The first one is very simple: we wish to make strong connections with our previous papers in which we have considered degeneracy of the electrons.  The second reason is that we consider it most appropriate to use the relationships such as equation (9) when we deal with relativistic electrons.  This choice reflects upon the mathematical preference of the authors who wish to use expressions as general as possible.

  We obtain 
\begin{equation}
< W(\omega) > d \omega \, = \, \frac{ \displaystyle{n_{e} n_{j} Z_{j}^{2} \, \alpha \, r_{0}^{2} \, \hbar c \lambda^{3} \, J^{-}(\lambda, \nu, u, Z_{j})}}{ \displaystyle{G_{0}^{-}(\lambda, \nu)}} \, d \omega \, , 
\end{equation}
\begin{eqnarray}
& & J^{-}(\lambda, \nu, u, Z_{j})  \nonumber  \\
& = & \int_{\lambda^{-1}+u}^{\infty} dx \, \frac{x^{2} - \lambda^{-2}}{ e^{x - \nu} \, + \, 1} \, \frac{x - u}{x} \left(1 - \frac{1}{e^{x-u-\nu}+1} \right) \nonumber \\
& \times &  \frac{1 - {\rm exp} \left[-2 \pi \alpha Z_{j} x \left(x^{2} - \lambda^{-2} \right)^{-1/2} \right] }{1 - {\rm exp} \left\{ -2 \pi \alpha Z_{j} (x - u) \left[ (x - u )^{2} - \lambda^{-2} \right]^{-1/2} \right\} }  \nonumber \\
& \times & \left( \frac{4}{3} \, - \, 2 (x - u) x \, \frac{ [ (x - u)^{2} -  \lambda^{-2} ] \, + \, ( x^{2} - \lambda^{-2})}{ [ (x - u)^{2} -  \lambda^{-2} ] \, ( x^{2} - \lambda^{-2})} \right. \nonumber  \\
& + & \lambda^{-2} \left\{ \frac{ \beta_{f} \, x}{ [ (x - u)^{2}  -  \lambda^{-2} ]^{3/2}}  +  \frac{ \beta_{i} \, (x - u)}{(x^{2} -  \lambda^{-2} )^{3/2}}  -  \frac{ \beta_{f} \, \beta_{i}}{ [(x - u)^{2}  - \lambda^{-2} ]^{1/2} (x^{2} - \lambda^{-2} )^{1/2}} \right\} 
 \nonumber \\
& + & L \, \left[ \frac{8}{3} \, \frac{(x - u) x}{ [(x - u)^{2}  - \lambda^{-2} ]^{1/2} (x^{2} - \lambda^{-2} )^{1/2}}  \right.  \nonumber  \\
&  & + \, \frac{u^{2}}{ [(x - u)^{2}  - \lambda^{-2} ]^{3/2} (x^{2} - \lambda^{-2} )^{3/2}} \left\{ (x - u)^{2}x^{2} \, + \,  [(x - u)^{2} - \lambda^{-2} ] (x^{2} - \lambda^{-2} ) \right\}  \nonumber   \\
& & + \, \frac{\lambda^{-2} u}{ 2 [(x - u)^{2}  - \lambda^{-2} ]^{1/2} (x^{2} - \lambda^{-2} )^{1/2} } \times \left\{ \frac{(x-u)x + (x^{2} - \lambda^{-2})}{ (x^{2} - \lambda^{-2} )^{3/2}} \beta_{i} \right.   \nonumber \\ 
&  & \left. \, \left. \, \left. \, - \, \frac{(x-u)x + [(x-u)^{2} - \lambda^{-2}]}{ [(x-u)^{2} - \lambda^{-2} ]^{3/2}} \beta_{f} \, + \, \frac{2u(x-u)x}{ [(x-u)^{2} - \lambda^{-2} ] \, (x^{2}-\lambda^{-2})} \, \right\} \, \, \, \, \right] \, \, \, \, \right)  \, ,
\end{eqnarray}
\begin{eqnarray}
\beta_{f} & = & 2 \, \, {\rm ln} \frac{(x-u) + [ (x-u)^{2} - \lambda^{-2} ]^{1/2}}{\lambda^{-1}}  \, ,  \\
\beta_{i} & = & 2 \, \, {\rm ln} \frac{x + (x^{2} - \lambda^{-2})^{1/2}} {\lambda^{-1}}  \, ,  \\
L & = & 2 \, \, {\rm ln} \frac{(x-u)x + [ (x-u)^{2} - \lambda^{-2}]^{1/2}  (x^{2}-\lambda^{-2})^{1/2} - \lambda^{-2}}{\lambda^{-1} u}  \, ,  \\
u & = & \frac{\hbar \omega}{k_{B}T}  \, .
\end{eqnarray}
By doing a transformation $x^{\prime} = x - u$, one finds that the formula (16) is related to $I^{-}(\lambda, \nu, u, Z_{j})$ of Itoh, Nakagawa, \& Kohyama (1985) (their equation (12)) by
\begin{equation}
J^{-}(\lambda, \nu, u, Z_{j}) \, = \, e^{-u} \, I^{-}(\lambda, \nu, u, Z_{j}) \, .
\end{equation}

  We define the temperature-averaged relativistic Gaunt factor $g_{Z_{j}}$ for the thermal bremsstrahlung emissivity by
\begin{equation}
< W(\omega) > d \omega \, = \, g_{Z_{j}} \, < W(\omega) >_{K} d \omega  \, , 
\end{equation}
where
\begin{eqnarray}
< W(\omega) >_{K} d \omega  & = & \frac{2^{5} \pi e^{6}}{3 h m c^{3}} \, n_{e} n_{j} Z_{j}^{2} \, \left( \frac{2 \pi k_{B} T}{3 m} \right)^{1/2} \, e^{-u} \, \frac{\hbar}{k_{B} T} \, d \omega  \nonumber  \\
& = & 1.426 \times 10^{-27} \left[ n_{e}({\rm cm^{-3}}) \right] \left[n_{j}({\rm cm^{-3}}) \right] Z_{j}^{2} \left[ T({\rm K}) \right]^{1/2} \nonumber  \\
& \times & e^{-u} \, du  \, \hspace{1.0cm} \, {\rm erg \, \, s^{-1} \, cm^{-3}}  \,  .
\end{eqnarray}
In the above we have followed the expression of the original paper by Karzas \& Latter (1961) as closely as possible.
Therefore we obtain
\begin{equation}
g_{Z_{j}} \, = \, \frac{3 \sqrt{6}}{32 \sqrt{\pi}} \, \lambda^{7/2} \, e^{u} \, \frac{ J^{-}(\lambda, \nu, u, Z_{j})}{G_{0}^{-}(\lambda, \nu)}  \, .
\end{equation}
We note that the formula (24) is identical to equation (21) of Itoh, Kohyama, \& Nakagawa (1985) with the positron contributions omitted, because one has $e^{u} \, J^{-}(\lambda, \nu, u, Z_{j})$ = $I^{-}(\lambda, \nu, u, Z_{j})$.

  In this paper we will also calculate the exact nonrelativistic energy-dependent Gaunt factor for bremsstrahlung, and then use this for the calculation of the thermally-averaged Gaunt factor.  The formulation is similar to the one presented in the paper by Nakagawa, Kohyama, \& Itoh (1987).  However, for the sake of completeness we will present it here.  According to Karzas \& Latter (1961), the exact nonrelativistic inverse bremsstrahlung Gaunt factor is given by
\begin{eqnarray}
g & = & \frac{2 \sqrt{3}}{\pi \eta_{i} \eta_{f}} \left[ \left( \eta_{i}^{2} + \eta_{f}^{2} + 2 \eta_{i}^{2} \eta_{f}^{2} \right) \, I_{0} - 2 \eta_{i} \eta_{f} \left(1 + \eta_{i}^{2} \right)^{1/2} 
\left(1 + \eta_{f}^{2} \right)^{1/2}  \, I_{1} \, \right] \, I_{0} \, , \\
\eta_{i}^{2} & = & \frac{Z_{j}^{2} {\rm R_{y}}}{\epsilon_{i}} \, \, , \hspace{0.5cm}  \eta_{f}^{2} \, \, \, = \, \, \, \frac{Z_{j}^{2} {\rm R_{y}}}{\epsilon_{f}} \,  \, ,  \\
I_{l} & = & \frac{1}{4} \left[ \frac{4 k_{i} k_{f}}{(k_{i} - k_{f})^{2}} \right]^{l+1} {\rm exp} \left( \frac{ \pi \mid \eta_{i} - \eta_{f} \mid }{2} \right) \frac{ \mid \Gamma (l+1+i \eta_{i}) \Gamma(l+1+i \eta_{f}) \mid}{\Gamma(2 l + 2)} \, G_{l}  \, ,  \\
G_{l} & = & \left| \frac{k_{f} - k_{i}}{k_{f} + k_{i}} \right|^{i \eta_{i} + i \eta_{f}} \, _{2}F_{1} \left[ l+1- i \eta_{f}, l+1-i \eta_{i}; 2l+2; - \frac{4 k_{i} k_{f}}{(k_{i} - k_{f})^{2}} \right] \, , 
\end{eqnarray}
\begin{equation}
 _{2}F_{1} (a,b;c;x) \, \equiv \, \sum_{m=0}^{ \infty} \frac{ \Gamma(a+m) \Gamma(b+m) \Gamma(c)}{ \Gamma(c+m) \Gamma(a) \Gamma(b) }  \, 
\frac{x^{m}}{m! }   \, .
\end{equation}
In the above $\epsilon_{i}$ and $\epsilon_{f}$ are the kinetic energies of the initial electron and final electron for the inverse bremsstrahlung, $k_{i}$ and $k_{f}$ are their wave numbers, and $_{2}F_{1}(a,b;c;x)$ is the hypergeometric function.  Karzas \& Latter (1961) have given a power series expansion method for the practical evaluation of $G_{l}$.  We have employed their method in the present calculation.

  The thermally-averaged Gaunt factor for the inverse bremsstrahlung by nonrelativistic electrons is written as 
\begin{eqnarray}
g_{NR}(u, \gamma^{2}) & = & \frac{ \pi^{1/2}}{2} \frac{ \displaystyle{ \int_{0}^{ \infty} d \epsilon_{i} \, g \, f(\epsilon_{i}) \left[ 1 - f(\epsilon_{f}) \right]}}
{ \displaystyle{\beta^{1/2} \int_{0}^{ \infty} d \epsilon_{i} \, \epsilon_{i}^{1/2} f(\epsilon_{i})} }  \, ,  \\
f( \epsilon_{i} ) & = & \left[ {\rm exp} ( \beta \epsilon_{i} - \eta) + 1 \right]^{-1}  \,  ,  \, \, \, \beta \, \equiv \, (k_{B}T)^{-1}  \,  , \\
\gamma^{2} & = & \frac{Z_{j}^{2} {\rm R_{y}}}{k_{B}T}  \, = \, Z_{j}^{2} \, \frac{1.579 \times 10^{5} {\rm K}}{T} \, .
\end{eqnarray}
Therefore, the nonrelativistic thermal bremsstrahlung emissivity is 
given by 
\begin{eqnarray}
< W(\omega) >_{NR} d \omega  & = & 1.426 \times 10^{-27} \, g_{NR}(u, \gamma^{2}) \, \left[ n_{e}({\rm cm^{-3}}) \right] \left[n_{j}({\rm cm^{-3}}) \right] Z_{j}^{2} \, \\  \nonumber
& \times &  \left[ T({\rm K}) \right]^{1/2} e^{-u} \, du  \, \hspace{1.0cm} \, {\rm erg \, \, s^{-1} \, cm^{-3}}  \,  .
\end{eqnarray}
The thermally-averaged nonrelativistic Gaunt factor $g_{NR}(u, \gamma^{2})$ for the inverse bremsstrahlung has been calculated and tabulated by Nakagawa, Kohyama, \& Itoh (1987), Itoh, Kojo, \& Nakagawa (1990), and by Itoh et al. (1997), for high-temperature, high-density plasmas.  Carson (1988) calculated the thermally-averaged nonrelativistic Gaunt factor for the inverse bremsstrahlung for non-degenerate electrons following the method of Karzas \& Latter (1961).

  In order to assess the accuracy of the various approximations, we have also calculated the nonrelativistic Gaunt factors with the Born approximation corrected by the Elwert factor.  It is given as
\begin{equation}
g_{NRE}  \, = \, \frac{\sqrt{3}}{\pi} {\rm ln} \left| \frac{p_{f} + p_{i}}{p_{f} - p_{i}} \right| \, \frac{\eta_{f}}{\eta_{i}} \frac{1 - {\rm exp} ( - 2 \pi \eta_{i})}{1 - {\rm exp} ( - 2 \pi \eta_{f})}  \, .
\end{equation}
The thermally-averaged nonrelativistic Gaunt factor in the Elwert approximation is obtained by inserting eq.\ (34) into eq.\ (30).

\section{NUMERICAL RESULTS}

  We have carried out the numerical calculations of the thermally averaged relativistic and nonrelativistic Gaunt factors for the thermal bremsstrahlung.  In the present paper we are interested in the high-temperature and low-density regime which is relevant to the hot gas in the clusters of galaxies.  The electron plasma in this regime is extremely nondegenerate, namely $\eta \approx - \infty$.  It is clear from Fig.\ 1 of Nozawa, Itoh, \& Kohyama (1998) that the relativistic Gaunt factor is independent of $\eta$ for negatively large values ($\eta \le -10$).  Therefore we adopt  $\eta=-70$ as the case of $\eta \approx -\infty$ throughtout this paper.

  In Fig.\ 1 we have plotted the thermally averaged relativistic Gaunt factor as a function of the photon energy $u$ for $Z_{j}$=26, $\eta=-\infty$ and the electron temperatures $T=10^{5}$K, 10$^{6}$K, 10$^{7}$K, 10$^{8}$K, 10$^{8.5}$K and 10$^{9}$K.  The relativistic effect becomes very important in the high $T$, large-$u$ region.  For the typical plasmas in the clusters of galaxies of $T=10^{8}$K the effect is large for the photon energies log$_{10}u \ge 2$.

  In Figs.\ 2 and 3 we have plotted the thermally averaged relativistic and nonrelativistic Gaunt factors for $\eta=-\infty$, log$_{10}u$=0 and 1 as a function of the temperature parameter $\gamma^{2}$ defined by eq.\ (32).  The dashed curve is the nonrelativistic exact calculation of eq.\ (30), where the thermally averaged Gaunt factor does not depend on $Z_{j}$ and $T$ separately, but on the combination $Z_{j}^{2}/T$.  The solid curves from left to right are the relativistic cases for $Z_{j}$=10, 12, 14, 16, and 26, respectively.

  The numerical results are also presented in Tables 1 and 2 for $\eta=-\infty$, $10^{-3.0} \leq \gamma^{2} \leq 10^{2.5}$, and $10^{-4.0} \leq u \leq 10^{3.0}$.  The first, second, third, fourth, fifth, and sixth entries correspond to the thermally averaged relativistic Gaunt factors of neon, magnesium, silicon, sulphur, iron, and the thermally averaged exact nonrelativistic Gaunt factor.  We wish to point out the following fact: for 0.0 $\leq$ log$_{10}u \leq 1.0$ and for 10 $\leq Z_{j} \leq 26$ there exists a certain range of $\gamma^{2}$ where the relativistic Elwert result almost coincides with the nonrelativistic exact result.  Let us consider the case of $Z_{j}=26$.  From Table 1 we find that for log$_{10}u$=0.0 both results show an excellent agreement (0.1\% accuracy) for log$_{10}\gamma^{2}=-0.5$.  For log$_{10}u$=1.0 both results show a very good agreement (0.4\% accuracy) for log$_{10}\gamma^{2}$=1.0.  This fact guarantees the accuracy of the relativistic Elwert calculation.  At high temperatures (small values of $\gamma^{2}$) the discrepancy is caused by the insufficiency of the nonrelativistic approximation.  The discrepancy becomes larger as $Z_{j}$ increases.  At low temperatures (large values of $\gamma^{2}$) the Coulomb distortion of the wave function becomes very large and the Elwert approximation becomes less accurate.  The accuracy of the numerical calculation of the exact nonrelativistic Gaunt factor for low--temperature cases is about 0.1\%.

  In order to assess the accuracy of the various approximations, we have also calculated the thermally averaged nonrelativistic Gaunt factors using the Elwert approximation.  We have found that the nonrelativistic Gaunt factor with the Elwert approximation coincides with the nonrelativistic exact Gaunt factor for $\gamma^{2} \leq 10^{-1.0}$ with an accuracy of better than 0.2\% for the photons of moderate energy (say log$_{10}u \sim 0$), thereby proving the excellence of the Elwert approximation at high temperatures.  It is also found that the nonrelativistic Gaunt factor with the Elwert approximation coincides with the relativistic Gaunt factor with the Elwert approximation at low temperatures, as it should.  For the cases of heavy elements, the Elwert approximation has a lower accuracy compared with the cases of hydrogen and helium.  Agreement of the results of the calculations with different methods proves the accuracy of the present calculations.

  For clusters of galaxies, the relativistic correction is not too large.  Take an example of $Z_{j}=26$ and $\gamma^{2}=1$ ($T=1.067 \times 10^{8}$K).  From Table 1 one finds $g_{R}/g_{NR}$=1.019 for log$_{10}u$=0.0 and $g_{R}/g_{NR}$=0.977 for log$_{10}u$=1.0.  Therefore, the relativistic correction in the case of Fe is on the order of 2\% for clusters of galaxies.

\section{CONCLUDING REMARKS}

  We have calculated the Gaunt factor for the thermal bremsstrahlung in high-temperature plasmas for the cases of $Z$ = 10, 12, 14, 16, 26 by using the accurate relativistic cross section, and have compared the result with the Gaunt factor derived by using Sommerfeld's exact nonrelativistic cross section.  Significant deviations from the nonrelativistic results have been found for high temperature cases.

  We have presented the results in the form of extensive tables.  The accuracy of the Elwert approximation is better than 0.2\% for $\gamma^{2} \leq 10^{-1}$, log$_{10}u \sim 0$.  The overall accuracy of the present calculation is generally about 0.4\%.  The present paper will be useful to analyze the relativistic effects for the thermal bremsstrahlung in the high-temperature plasmas which exist in the clusters of galaxies. The present paper covers a much wider temperature range than that of the intracluster plasma for the sake of completeness.

\newpage


\references{} 
\reference{} Arnaud, K. A., Mushotzky, R. F., Ezawa, H., Fukazawa, Y., Ohashi, T., Bautz, M. W., Crewe, G. B., Gendreau, K. C., Yamashita, K., Kamata, Y., \& Akimoto, F. 1994, ApJ, 436, L67
\reference{} Bethe, H. A., \& Heitler, W. 1934, Proc. Roy. Soc. London, A146, 83
\reference{} Carson, T. R. 1988, A \& A, 189, 319
\reference{} Elwert, G. 1939, Ann. d. Physik, 34, 178
\reference{} Elwert, G., \& Haug, E. 1969, Phys. Rev., 183, 90.
\reference{} Gould, R. J. 1980, ApJ, 238, 1026
\reference{} Holzapfel, W. L. et al. 1997, ApJ, 480, 449
\reference{} Itoh, N., Kojo, K., \& Nakagawa, M. 1990, ApJS, 74, 291
\reference{} Itoh, N., Kuwashima, F., Ichihashi, K., \& Mutoh, H. 1991, ApJ, 382, 636
\reference{} Itoh, N., Nakagawa, M., \& Kohyama, Y. 1985, ApJ, 294, 17
\reference{} Itoh, N., et al. 1997, in AAS CD-ROM Series, Astrophysics on Disc, Vol. 9 (Washington: AAS)
\reference{} Karzas, W. J., \& Latter, R. 1961, ApJS, 6, 167
\reference{} Lee, C. M., Kissel, L., Pratt, R. H., \& Tseng, H. K. 1976, Phys. Rev., A13, 1714
\reference{} Markevitch, M., Mushotzky, R., Inoue, H., Yamashita, K., Furuzawa, A., \& Tawara, Y. 1996, ApJ, 456, 437
\reference{} Markevitch, M., Yamashita, K., Furuzawa, A., \& Tawara, Y. 1994, ApJ, 436, L71
\reference{} Nakagawa, M., Kohyama, Y., \& Itoh, N. 1987, ApJS, 63, 661
\reference{} Nozawa, S., Itoh, N., \& Kohyama, Y. 1998, ApJ, submitted
\reference{} Pratt, R. H., \& Tseng, H. K. 1975, Phys. Rev., A11, 1797
\reference{} Rephaeli. Y., \& Yankovitch, D. 1997, ApJ, 481, L55
\reference{} Tseng, H. K., \& Pratt, R. H. 1971, Phys. Rev., A3, 100


\newpage

\centerline{\bf \large Figure Captions}

\begin{itemize}

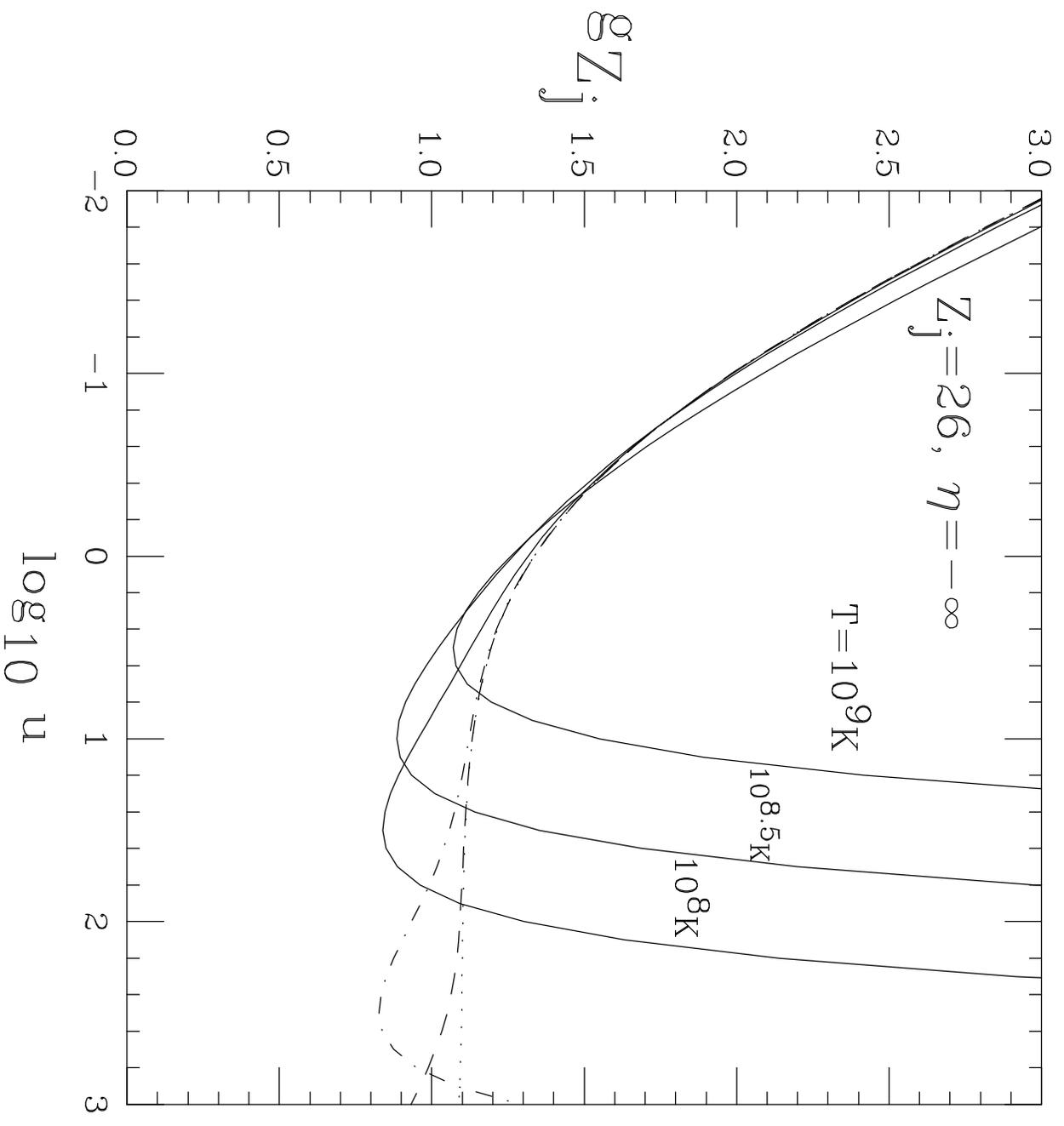
\item Fig.\ 1. Thermally averaged relativistic Gaunt factor for $Z_{j}=26$, $\eta=-\infty$.  The dotted curve corresponds to $T=10^{5}$K.  The dashed curve corresponds to $T=10^{6}$K.  The dash-dotted curve corresponds to $T=10^{7}$K.  The solid curves from right to left correspond to $T=10^{8}$K, $10^{8.5}$K and $10^{9}$K, respectively.

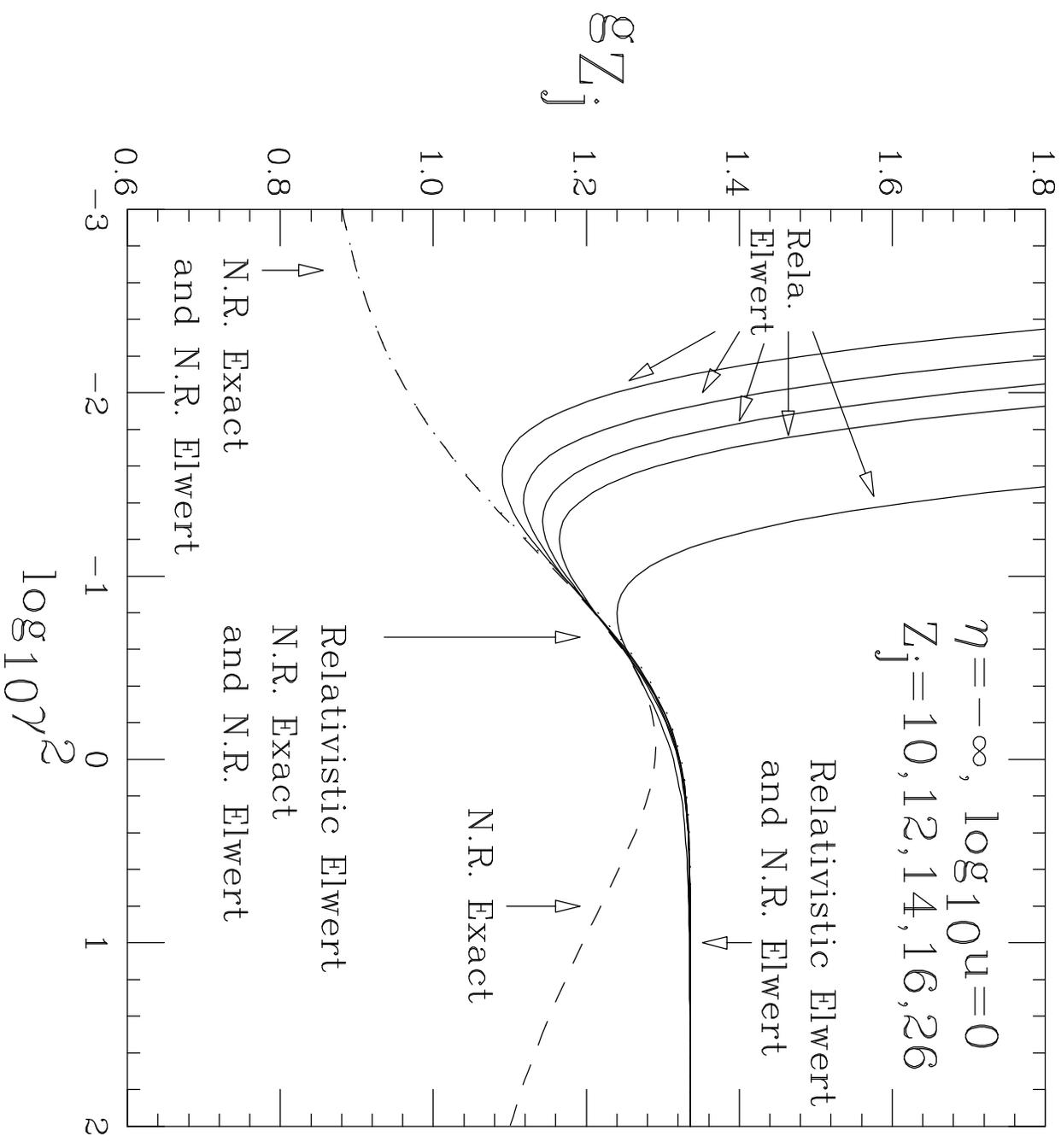
\item Fig.\ 2. Thermally averaged Gaunt factors for $\eta=-\infty$, log$_{10}u=0$.  The dashed curve corresponds to the nonrelativistic Gaunt factor (N.R. exact).  The solid curves from left to right correspond to the relativistic Gaunt factors for $Z_{j}$=10,12,14,16, and 26, respectively.  The nonrelativistic Gaunt factor in the Elwert approximation is also displayed as the dotted curve, but it is indistinguishable from other curves.

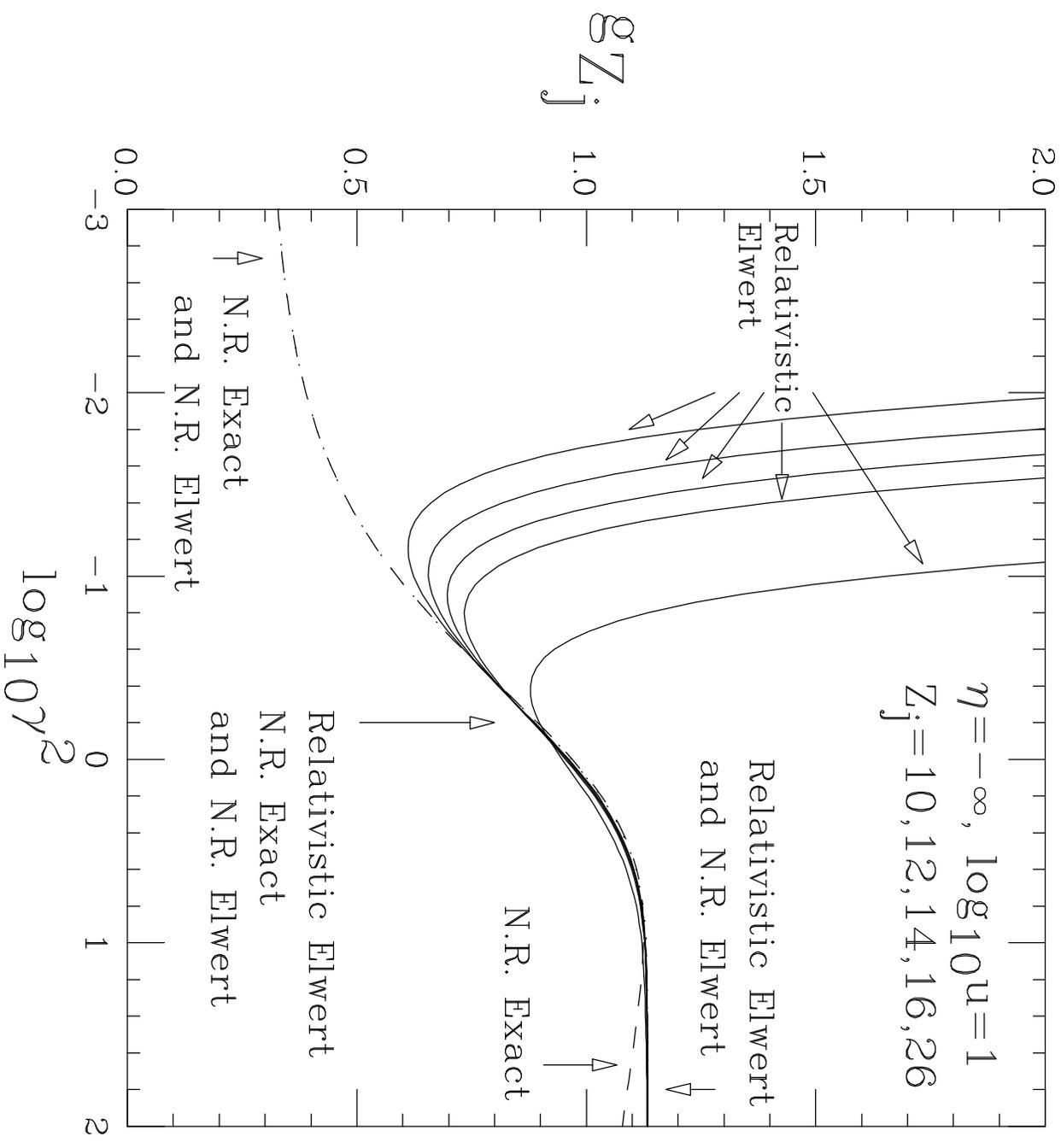
\item Fig.\ 3. Same as for Fig.\ 2 but for $\eta=-\infty$, log$_{10}u=1$.

\end{itemize}

\end{document}